\documentclass[aps,prc,twocolumn,superscriptaddress]{revtex4-1}

\usepackage{amsmath,amssymb,graphicx}
\usepackage{url}
\usepackage[colorlinks,urlcolor=blue]{hyperref}

\begin{document}

\title{A background-free signal of  jet-induced diffusion wake in quark-gluon plasma}

\author{Zhong Yang}
\email[]{zhong.yang@vanderbilt.edu}
\affiliation{Central China Center for Nuclear Theory \& Institute of Particle Physics, Central China Normal University, Wuhan 430079, China}
\affiliation{Department of Physics and Astronomy, Vanderbilt University, Nashville, TN}

\author{Xin-Nian Wang}
\email[]{xnwang@ccnu.edu.cn}
\affiliation{Key Laboratory of Quark and Lepton Physics (MOE) \& Institute of Particle Physics, Central China Normal University, Wuhan 430079, China}
\affiliation{Institut f\"{u}r Theoretische Physik, Johann Wolfgang Goethe–Universit\"{a}t, Max-von-Laue-Str.~1, D-60438 Frankfurt am Main, Germany}

\begin{abstract}
Rapidity asymmetry of jet-hadron correlation has been proposed as a robust signal of dijet‐induced diffusion wake in quark-gluon plasma in high‐energy heavy-ion collisions. We generalize this observable to other jet configurations such as $\gamma/Z^0$-jets and propose a new method to compute the rapidity asymmetry that is free of background. In this new method, the rapidity of the trigger is fixed or restricted to a symmetrical range while the rapidity of the associated jet is varied. The rapidity asymmetry is defined as the difference between the hadron rapidity distribution in the opposite azimuthal direction of the jet (or the same direction of the trigger) with different associated jet rapidities. Since the background in this hadron rapidity distribution with different associated jet rapidity is identical, it will be completely canceled in the rapidity asymmetry.  
Such background-free rapidity asymmetry caused by jet-induced diffusion wake is demonstrated in CoLBT-hydro simulations of dijets and $\gamma$-jets in Pb+Pb collisions at the LHC, which weakens as hadron $p_T$ increases, and subtraction of a p+p baseline has a negligible effect. 

\end{abstract}
\pacs{}

\maketitle

\section{Introduction}

Jet-induced diffusion wake is part of the medium response~\cite{Casalderrey-Solana:2004fdk, Stoecker:2004qu, Ruppert:2005uz, Chakraborty:2006md,Gubser:2007ga, Gubser:2007ni,Neufeld:2008fi,Betz:2008ka,Qin:2009uh, Li:2010ts, Bouras:2012mh, Yan:2017rku, Casalderrey-Solana:2020rsj, Cao:2020wlm, Mehtar-Tani:2022zwf}
to the interaction between a propagating jet and the Quark–Gluon Plasma (QGP)~\cite{Bjorken:1982tu, Thoma:1990fm, Braaten:1991we, Gyulassy:1993hr, Baier:1996kr, Zakharov:1996fv, Gyulassy:1999zd, Wiedemann:2000za, Wang:2001ifa, Arnold:2002ja, Djordjevic:2006tw, Qin:2007rn} in high-energy heavy-ion collisions. Experimental measurements of such jet-induced medium response can help shed lights on the jet quenching mechanism, thermalization and properties of QGP. With many theoretical and phenomenological studies~\cite{Ma:2010dv,Betz:2010qh,Tachibana:2014lja,Casalderrey-Solana:2016jvj,Tachibana:2017syd,Chen:2017zte, Pablos:2019ngg,Chen:2020tbl,Tachibana:2020mtb,Yang:2022yfr,Du:2022oaw,Yang:2023dwc,Xiao:2024ffk,Bossi:2024qho,Barata:2024ieg} of the consequences of the jet-induced medium response, experimental measurements of medium modification of jet substructures and correlations \cite{CMS:2018jco,CMS:2018mqn, ATLAS:2020wmg, CMS:2021otx,PHENIX:2024twd} so far still have not identified unambiguously effects of the medium response since they often entangle with the effects of medium-induced gluon radiation in the direction of the propagating jets. Though the depletion of soft hadron yield by the diffusion wake is unique, the smallness of the signal can still be overwhelmed by other effects \cite{Chen:2017zte,Yang:2021qtl,CMS:2021otx} and is difficult to measure. Motivated by a proposal and model calculation  of the 2D jet-hadron correlation in rapidity and azimuthal angle\cite{Yang:2022nei}, in which diffusion wake manifests as a valley structure on top of a multiple-parton-interaction (MPI) ridge, ATLAS and CMS Collaboration have indeed observed recently the direct evidence of diffusion wake in jet–hadron correlation in $\gamma$–jet events~\cite{ATLAS:2024prm} and $Z^0$–hadron correlations in $Z^0$–jet events~\cite{CMS:2024fli,CMS:2025dua}, respectively. More recently, we have proposed a new observable to search for the signal of diffusion wake in dijet events. This novel approach not only overcomes the limited availability of Z/$\gamma$-jet events in earlier measurements but also relaxes stringent detector requirements and background subtractions, thereby enabling experimental measurements at both RHIC and LHC with increased statistics and less background contamination.

The key challenge in searching for the jet-induced diffusion wake lies in distinguishing particles produced by the jet-induced medium response from those originating from jet hadronization and the underlying QGP background.  In theoretical calculations, the separation is straightforward. One can look for signals away from jets and unambiguously tag particles as either jet‐induced or background, yielding a clean signal of the diffusion‐wake. In experimental measurements, however, particles from medium response and the background are both soft with signal to background ratio in the order of a small fraction of a percentage. Separating such a small signal from a huge background is extremely challenging if not impossible. To avoid the problem of background subtraction, ATLAS only measures the ratio of jet-hadron correlation in each event to that in mixed-events~\cite{ATLAS:2024prm}, leading to a very weak signal of diffusion wake $\sim 0.5 \%$, consistent with our model calculation. The CMS Collaboration~\cite{CMS:2025dua} uses the  average distribution of mixed events to model the QGP background, enabling the extraction of a clear signal of diffusion wake. However, because of particle‐number conservation, this mixed-event method also produces a negative $Z^0$–hadron correlation in the $Z$ direction in p+p collisions. Though the shape of such a negative $Z^0$–hadron correlation in p+p due to event-mixing is very different from that in Pb+Pb collisions, it still casts a shadow on the genuineness of the observed signal of the diffusion wake. It is, therefore, crucial to devise a new observable that can be used to identify the diffusion-wake signal without performing background subtraction. 

Our analysis of diffusion wakes in dijet events shows that the jet–hadron rapidity correlations become strikingly asymmetric when we compare events with no/small and large dijet rapidity gaps~\cite{Yang:2025dqu}. This asymmetry offers unambiguous evidence for the presence of the diffusion wake in dijet events and it is close to background-free because the QGP background should be in principle unaffected by the dijet rapidity gap.  
Since the rapidity of the trigger (leading) jet could be different for different dijet rapidity gap, the underlying jet production cross section would be different, leading to a slightly different hadron rapidity distribution in the direction of the trigger where one is looking for the diffusion wake signal. Consequently, the asymmetry in the jet-hadron correlation is not entirely free of the background which still needs to be subtracted.
Furthermore, the jet-hadron correlation involves jets with non-zero rapidity, a rapidity-dependent  acceptance along the longitudinal direction will make measuring the asymmetry difficult.

To overcome the issues described above, one should avoid the background’s dependence on jet rapidity. A practical solution is to replace the conventional jet–hadron rapidity correlation with an analysis of the hadron rapidity distribution alone for given or fixed range of the trigger rapidity, either the leading jet in dijet events or $\gamma/Z^0$ in $\gamma/Z^0$-jet events. In this paper, we show the rapidity asymmetry,  defined as the difference between hadron rapidity distributions with different associated jet rapidities in $\gamma$-jet and dijet events, is a clear and unambiguous signal of the diffusion wake and is genuinely background-free. Such measurement is also straightforward since the raw hadron rapidity distribution is easy to measure experimentally.

\section{Rapidity asymmetry in $\gamma$-jets}

We use the Coupled Linear Boltzmann Transport hydrodynamic (CoLBT-hydro) model~\cite{Chen:2020tbl, Chen:2017zte, Zhao:2021vmu} to simulate the $\gamma$-jet production and transport in the QGP medium in 0-10$\%$ Pb+Pb collisions at $\sqrt{s}=5.02$ TeV. The initial condition for the bulk matter production in CoLBT-hydro is provided by the Trento model~\cite{Moreland:2014oya} which is also used to sample the initial transverse position for the hard processes. The initial $\gamma$-jet configuration is generated by PYTHIA8~\cite{Sjostrand:2007gs, Sjostrand:2014zea, Bierlich:2022pfr} with kinetic cuts, $p_T^\gamma>$ 100 GeV/c and $|\eta_\gamma|<1.44$. Final jets are reconstructed using FASTJET~\cite{Cacciari:2011ma} with anti-$k_T$ algorithm. We choose R=0.4 and require $p_T^{\rm{jet}}>$ 50 GeV/c and $|\eta_{\rm{jet}}|<1.6$.

Since the diffusion wake depletes soft hadrons in the opposite direction of the jet, we first suppress contributions from jet fragmentation by selecting final-state hadrons with $\Delta \phi^{\rm{jet}, h}=|\phi_h-\phi_{\rm{jet}}|>\pi/2$. Next, we divide $\gamma$-jet events into two rapidity classes: $|\eta_{\rm{jet}}|>$ 0.8 and $|\eta_{\rm{jet}}|<$ 0.8. Our previous studies show that the “valley” in the hadron rapidity distribution produced by the diffusion wake is tied to the rapidity of the propagating jet that induces the medium response. By separating events this way, the valleys in the two cases appear at different positions along the rapidity direction, leading to a clear rapidity asymmetry -- the difference between the hadron rapidity distributions in the two cases, as a consequence of the jet-induced diffusion wakes. As a convention in our analysis, we require $\eta_{\rm{jet}} > 0.0$, so that the signal of the diffusion wake appears in the positive rapidity region.
\begin{figure}[h!]
\centering
	\includegraphics[width=0.48\textwidth]{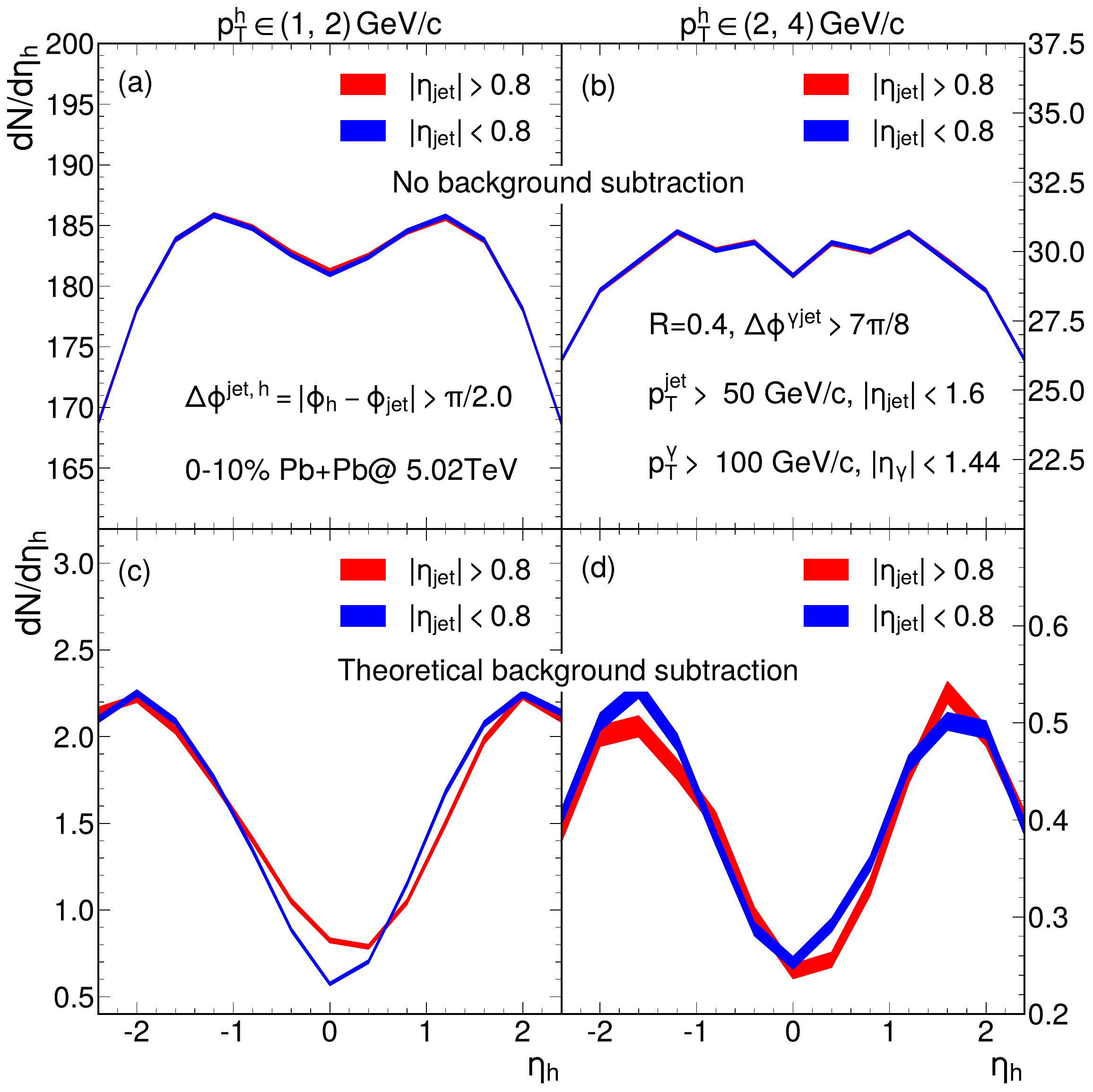}
	\caption{Rapidity distributions of soft hadrons in the opposite direction of jet within $p_T^h\in(1,2)$ GeV/c (a, c) and $p_T^h\in(2,4)$ GeV/c (b, d) in 0-10$\%$ Pb+Pb collisions at 5.02 TeV. The upper panel includes hydro background and lower panel has the hydro background subtracted. The red band is from the case of $|\eta_{\rm{jet}}|>$ 0.8 and blue band is from the case of $|\eta_{\rm{jet}}|<$ 0.8, respectively.}
	\label{jhr}
\end{figure}

We first show in Figs.~\ref{jhr}~(a) and (b) the total hadron rapidity distributions in the opposite azimuthal direction of the jet from the CoLBT-hydro simulations with two jet rapidity ranges (red and blue bands). In both transverse momentum intervals, $p_T^h = 1\text{–}2\;\text{GeV}/c$ and $p_T^h = 2\text{–}4\;\text{GeV}/c$, we observe no apparent differences between the case for $|\eta_{\text{jet}}| > 0.8$ (red bands) and the case for $|\eta_{\text{jet}}| < 0.8$ (blue bands), because the total distributions are dominated by hadrons from the hadronization of the bulk QGP background. The effect of the diffusion wake in the order of a fraction of a percentage is difficult to be recognizable. This is consistent with the ATLAS measurement~\cite{ATLAS:2024prm}.
\begin{figure}[h!]
\centering
	\includegraphics[width=0.48\textwidth]{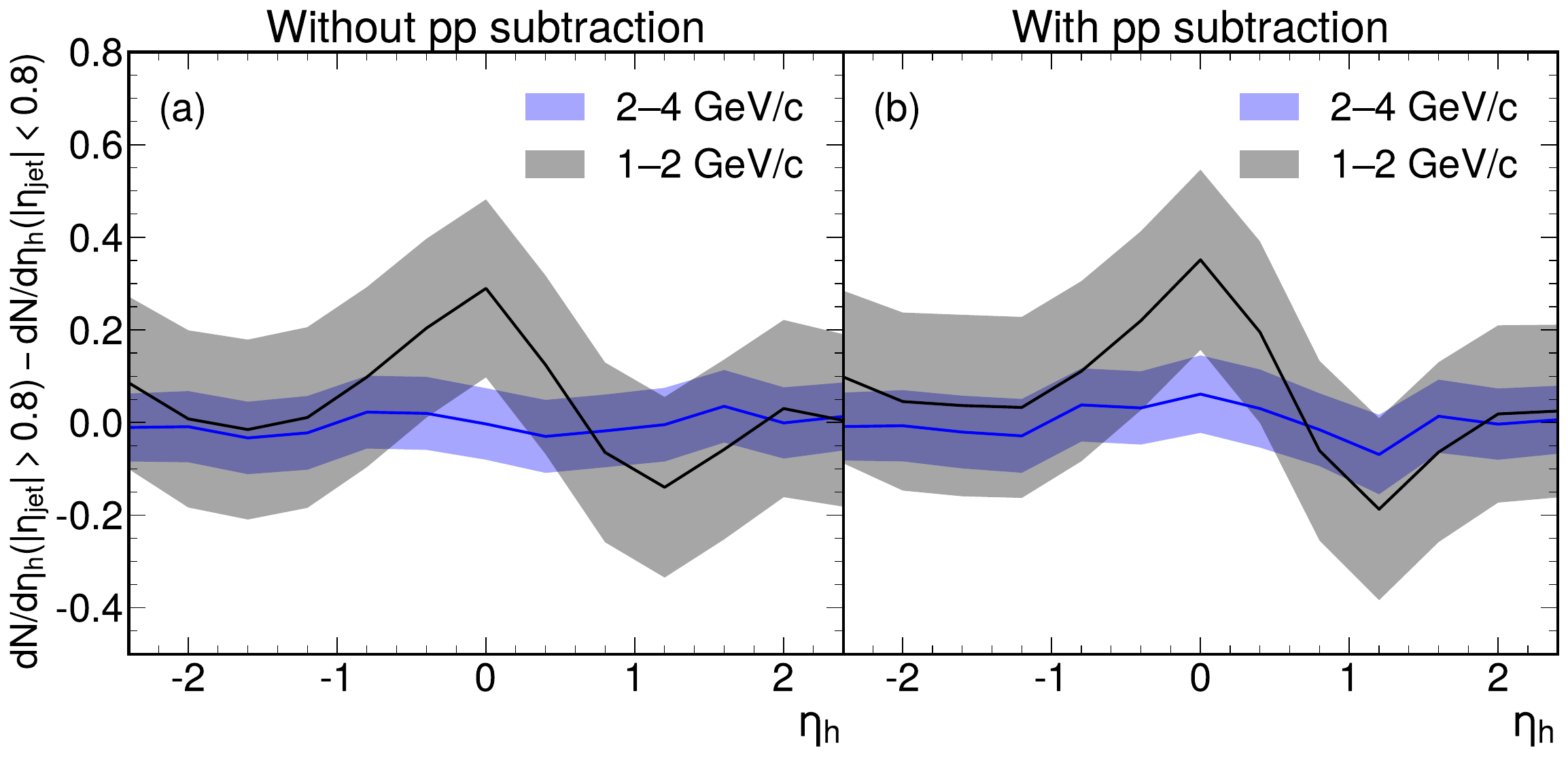}
	\caption{Rapidity asymmetry between soft hadron rapidity distributions  in the opposite direction of the jet ($\Delta\phi_{\rm{jet},h}>\pi/2.0$) in two different classes of $\gamma$-jets:  (1) $|\eta_{\rm{jet}}|>$ 0.8 and (2) $|\eta_{\rm{jet}}|<$ 0.8 for $p_T^h\in(1,2)$ GeV/c (black lines with gray bands) and  $p_T^h\in(2,4)$ GeV/c (blue lines with blue bands). The left panel is for 0-10 $\%$ Pb+Pb collisions and right panel has the p+p baseline subtracted.}
	\label{subdw}
\end{figure}
The similar distributions in Figs.~\ref{jhr} (a) and (b) for different jet rapidity ranges also indicate that the QGP background in these two cases is nearly identical. In Figs.~\ref{jhr} (c) and (d), we subtract the theoretical background which is obtained by running the CoLBT-hydro model with the same initial conditions but without jets..
The valley in rapidity on top of the hadron distribution from MPI is caused by the diffusion wake as shown in the previous study \cite{Yang:2022nei}. One can clearly see that the position of the valley changes with the rapidity of the jet. 

To avoid background subtraction which is difficult in experiments, we propose to calculate the rapidity asymmetry as the difference between the two hadron rapidity distributions [in Figs.~\ref{jhr} (a) and (b)] for events with two different cases of jet rapidity as shown in Fig.~\ref{subdw}. The QGP background should be canceled in this asymmetry. In the left panel [Fig.~\ref{subdw} (a)], the black curve represents the asymmetry for soft‐hadrons with $1 < p_T^h < 2$ GeV/c. For jets with $|\eta_{\rm jet}|>0.8$, the diffusion wake induces a depletion of soft hadrons near $\eta_h\approx 1.0$, while for $|\eta_{\rm jet}|<0.8$ the depletion occurs around $\eta_h\approx0.0$, as seen in Figs.~\ref{jhr} (c) and (d). The difference between these two rapidity selections thus produces an asymmetric soft‐hadron rapidity distribution, an unambiguous signal of the diffusion wake in high-energy heavy-ion collisions without background subtraction. In $2 < p_T^h < 4$ GeV/c range (blue curve), the medium-response effect is small and  there is no clear distinction between the hadron rapidity distributions for different jet rapidity selections.  Therefore rapidity asymmetry is very small.

In the right panel of the figure [Fig.~\ref{subdw} (b)], we also subtract the corresponding hadron rapidity distributions in p+p collisions to obtain the pure medium modification. We find that the p+p baseline has a negligible effect, confirming that the asymmetric rapidity distribution is indeed caused by the jet-induced diffusion wake.

\begin{figure}[h!]
\centering
	\includegraphics[width=0.48\textwidth]{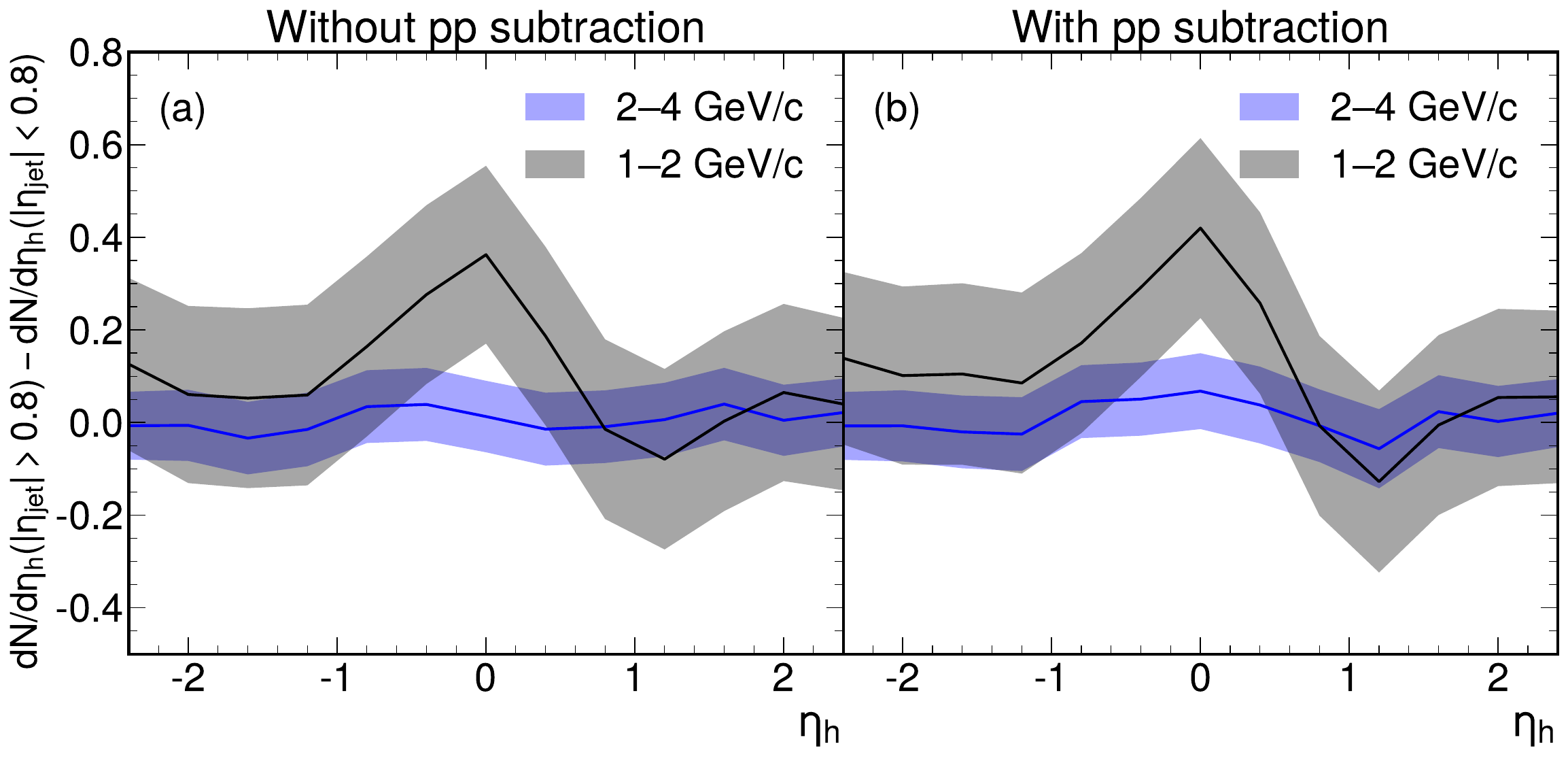}
	\caption{The same as Fig.~\ref{subdw} except that soft hadrons are selected with reference to $\gamma$, $\Delta \phi^{\gamma h}=|\phi_h-\phi_{\gamma}|<\pi/2.0$.}
	\label{subdwgam}
\end{figure}

We have also repeated the analysis using the photon direction as a reference in azimuthal angle for hadron selection. The results shown in Fig.~\ref{subdwgam} are about the same as in Fig.~\ref{subdw} where the opposite direction of the jet axis is used for reference.  Since photon and jet are not always perfectly back to back because of vacuum radiation and medium interactions,  the misalignment smears the diffusion-wake signal. As a result, the valley depth in Fig.~\ref{subdwgam} is smaller than in Fig.~\ref{subdw}. Nevertheless, whether the $\gamma$ or the jet is used as the reference for hadron selection, this background‐free observable reveals a clear signal of the diffusion wake induced by $\gamma$-jets. 

\section{ Rapidity asymmetry in dijets}

The above analysis can also be applied to dijet events in high-energy heavy-ion collisions. We again use the CoLBT-hydro model to simulate transport of dijets in QGP and calculate the rapidity asymmetry in central 0-10\% Pb+Pb collisions at $\sqrt{s}=5.02$ GeV. We select dijets with the transverse momentum $p_T^{\rm jet_1} > 120~\mathrm{GeV}/c$ for the leading jet and $p_T^{\rm jet_2} > 90~\mathrm{GeV}/c$ for the subleading jet. The rapidity cut of dijet is $|\eta_{\rm{jet}_1,\rm{jet}_2}|<1.6$. In addition, we impose $|\Delta\phi_{\rm jet_1,jet_2}| > \pi/2$ to ensure that the leading and subleading jets are approximately back‑to‑back. 

We treat the leading jet as a trigger, analogous to the $\gamma$ trigger in $\gamma$-jet events, and require $\Delta\phi_{\rm{jet}_1,h}<\pi/2.0$ to select particles on the away side of the subleading jet. The rapidity distribution of these soft hadrons should contain information about the diffusion wake from the subleading jet.  As a convention in the calculation, we require $\eta_{\rm{jet}_1}>0$ and inverse the rapidity when $\eta_{\rm{jet}_1}<0$. We divide events into two classes according to the relative positions of the two jets in rapidity: (1) $\eta_{\rm{jet}_1}\eta_{\rm{jet}_2}>0$, for two jets in the same hemisphere with small dijet rapidity gap and (2) $\eta_{\rm{jet}_1}\eta_{\rm{jet}_2}<0$, for two jets in the opposite hemisphere with large dijet rapidity gap. This strategy allows us to position the diffusion wake of the subleading jet in different rapidity regions in the two classes of events, 
while ensuring the soft-hadron enhancement induced by the leading jet happens in the same rapidity region. Consequently, in the rapidity asymmetry or the difference between the rapidity distributions of the selected soft hadrons in the two classes of events, not only the hydrodynamic background is canceled  but also soft hadrons from the wake front and medium-induced gluon radiation of the leading jet are suppressed.
\begin{figure}[h!]
\centering
	\includegraphics[width=0.48\textwidth]{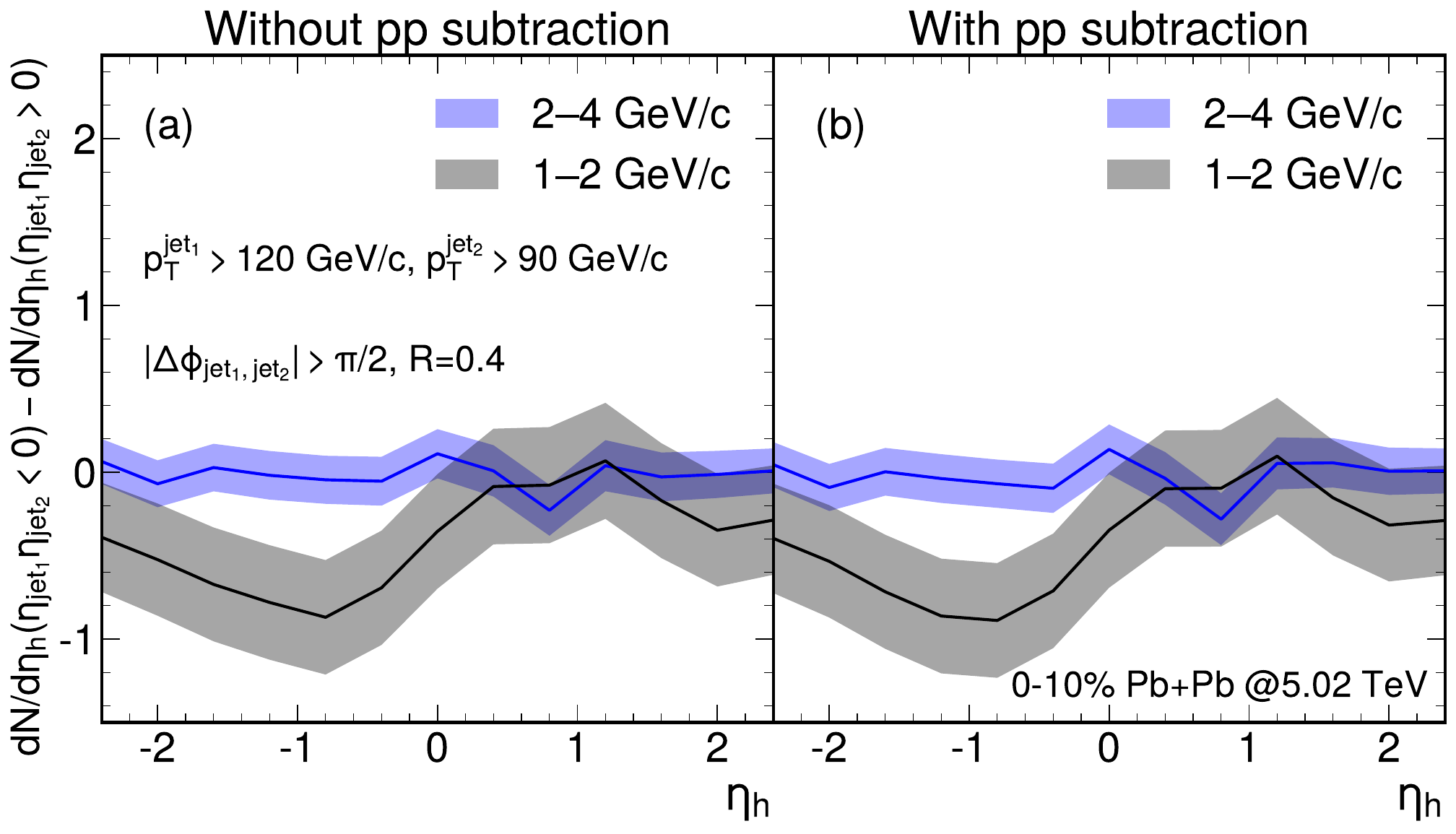}
	\caption{Rapidity asymmetry between soft hadron rapidity distributions  in the direction of the leading jet ($\Delta\phi_{\rm{jet}_1,h}<\pi/2.0$) in two different classes of dijets:  (1) $\eta_{\rm{jet}_1}\eta_{\rm{jet}_2}>0$ and (2) $\eta_{\rm{jet}_1}\eta_{\rm{jet}_2}<0$, for $p_T^h\in[1.0, 2.0]$ GeV/$c$ (black lines with gray bands) and $p_T^h\in[2.0, 4.0]$ GeV/c (blue lines with blue bands). The left panel is for 0-10 $\%$ Pb+Pb collisions and the right panel has the p+p baseline subtracted.}
	\label{subdwsin}
\end{figure}

As shown in Fig.~\ref{subdwsin}~(a), a similar rapidity asymmetry is observed for $p_T^h\in[1.0, 2.0]$ GeV/c. The dip near $\eta_h\approx -1.0$ mainly comes from the diffusion wake of the subleading jet in the $\eta_{\rm{jet}_1}\eta_{\rm{jet}_2}<0$ configuration, while the enhancement around $\eta_h\approx 1.0$ is due to the diffusion wake of the subleading jet in the $\eta_{\rm{jet}_1}\eta_{\rm{jet}_2}>0$ class. Such an asymmetric structure serves as a robust signal of the diffusion wake in dijet events without any background subtraction. This asymmetry is completely negligible for $p_T^h \in (2.0,4.0)$ GeV/$c$, as effects of the medium response diminish with increasing hadron $p_T$. In Fig.~\ref{subdwsin}~(b), we again show that subtracting the p+p baseline does not have any visible effect on rapidity asymmetry, consistent with our observations in $\gamma$‑jet events.

\section{Summary} 

In this work, we introduce a novel strategy to identify jet–induced diffusion wake in $\gamma$-jet in high-energy heavy-ion collisions. By examining the hadron rapidity distributions along the opposite direction of the associated jet and their dependence on hadron $p_T$ and jet rapidity, we find a distinctive asymmetry that emerges when contrasting distributions between different jet rapidity. Crucially, this observable is intrinsically free of uncorrelated background, offering a clean and experimentally accessible signal of the diffusion wake. 

We further validate this background‑free approach in dijet events, where it remains equally effective. This straightforward extension enables a systematic search for diffusion‑wake across different jet topologies in high‑energy heavy‑ion collisions, without the need for background subtraction, greatly facilitating the experimental exploration of diffusion wakes and jet-induced Mach-cone in high‑energy heavy‑ion collisions.

\section*{Acknowledgment} 

This work is supported in part by the Postdoctoral Fellowship Program and China Postdoctoral Science Foundation under Grant No. BX20240134 (ZY) and China Postdoctoral Science Foundation under Grant No. 2024M751059 (ZY), by NSFC under Grant No. 1193507 and by the Guangdong MPBAR with No. 2020B0301030008. X.-N.W. is also supported by the Alexander von Humboldt Foundation through the Humboldt Research Award. Computations in this study are performed at the NSC3/CCNU and NERSC under the award NP-ERCAP0032607.

\bibliography{Refsdw}
\end{document}